%% file: main.tex
\documentclass[10pt,a4paper]{article}

\input{style}

\title{Ensemble size effects on conditional reliability estimates: slope attenuation bias and correction methods}

\author[1]{Jonas Spaeth\,\orcidlink{0009-0006-4514-3787}}
\author[1]{Christopher D. Roberts\,\orcidlink{0000-0002-2958-6637}}

\affil[1]{European Centre for Medium-Range Weather Forecasts, Bonn, Germany and Reading, England}

\date{\normalsize PREPRINT\\[1.5ex]\today}

\begin{document}

\maketitle
\thispagestyle{empty}


\begin{abstract}
The goal of ensemble forecasting is to maximise sharpness subject to reliability.
Marginal reliability means that, over all cases, the ensemble is statistically consistent with reality: the ensemble mean is unbiased, the expected ensemble variance equals the expected mean-squared error of the ensemble mean, and the variance of the ensemble members matches the variance of the truth.
Equivalently, forecasts that assign probability $p$ to an event verify with relative frequency $p$.
However, climatological consistency is not sufficient for users acting on individual forecasts.
A natural extension is to assess reliability conditional on the forecast itself, by examining whether, on average, larger ensemble means imply larger observed values, larger spreads imply larger forecast errors, or higher probabilities imply higher event frequencies.
This motivates conditional reliability diagnostics such as reliability diagrams and spread-error relationships.

Here we show that conditional reliability diagnostics are systematically biased for finite ensemble sizes.
We present a unified framework for slope attenuation caused by finite-ensemble sampling noise, which affects conditional diagnostics for ensemble means, spreads, and probabilities.
Using synthetic forecasts that are perfectly reliable by construction, we isolate finite-ensemble effects.
We derive analytical expressions for the expected attenuation and propose practical estimators computable directly from ensemble data.

The framework is illustrated using 2-metre temperature sub-seasonal ensemble forecasts from ECMWF, where finite-ensemble slope attenuation substantially affects the spread-error relationship and tercile-based reliability diagrams.
These results demonstrate that attenuated conditional slopes should not be interpreted as evidence of forecast deficiencies unless finite-ensemble effects are explicitly taken into account.
\end{abstract}

\section{Introduction}
\label{sec:introduction}

Forecast reliability is a central aspect of forecast quality in probabilistic and ensemble prediction and refers to the statistical consistency between forecast distributions and verifying observations \parencite{gneitingProbabilisticForecastsCalibration2007}.
In a reliable ensemble system, observations are statistically indistinguishable from the individual ensemble members.
Forecast reliability is crucial for the meaningful interpretation and practical use of ensemble forecasts, as it ensures that forecast probabilities, uncertainties, and expected values can be interpreted at face value in decision-making contexts. Moreover, deficiencies in reliability generally also limit achievable probabilistic skill, since systematic inconsistencies between forecast distributions and observations degrade the performance of probabilistic scoring measures \parencite{robertsEnsembleReliabilitySignaltonoise2025}.
The term ``calibration'' is often used synonymously with ``reliability'' in the literature.

To formalize the concept of reliability for ensemble predictions, we adopt the classical Monte-Carlo interpretation of ensemble forecasts, in which both the ensemble members and the verifying observation are viewed as independent realizations from the same, generally unknown, population distribution \parencite[e.g.,][]{gneitingProbabilisticForecastsCalibration2007,wilksReliabilityRankHistogram2011}.
This implies consistency requirements for different characteristics of that distribution \parencite[see, e.g.,][]{brockerConceptExchangeabilityEnsemble2011,weigel2011verification}.
At the level of the first moment, reliability requires that the climatological mean of the ensemble mean matches the climatological mean of the observations, such that the forecast system is unbiased on average. 
At the level of the second moment, reliability implies two independent conditions: first, that the climatological ensemble spread is statistically consistent with the mean squared error of the ensemble mean \parencite{https://doi.org/10.1002/qj.49711448010,grimitMeasuringEnsembleSpread2007,hopsonAssessingEnsembleSpread2014}; and second, that the variance (over different cases) of the verifying observations matches the variance of individual ensemble members \parencite{johnsonReliabilityCalibrationEnsemble2009}.
Finally, for event-based forecasts, reliability implies that forecasts assigning probability $p$ to an event -- such as an intense cyclone, heavy precipitation, or a severe heat wave -- verify with relative frequency $p$, when averaged over all such forecasts \parencite{https://doi.org/10.2307/2346866,brockerIncreasingReliabilityReliability2007}.

In practice, forecast reliability is often assessed through comparisons of long-term, climatological average statistics derived from forecasts and observations.
This is sometimes referred to as \textit{marginal reliability} or \textit{climatological reliability}, as it considers averages over all forecast cases without conditioning on specific forecast instances.

From a user perspective, however, interest typically extends beyond climatological average behaviour.
Decisions are usually based on individual forecast instances, and users therefore need to know whether ensemble-derived quantities -- such as the ensemble mean, ensemble spread, or forecast probabilities -- can be trusted as indicators of expected outcomes, forecast uncertainty, or event likelihood in a given situation.
In particular, users may ask whether the mean of the observations is higher in cases where the ensemble mean indicates higher values, whether forecast errors tend to be larger in cases where the ensemble spread is larger, and whether events occur more frequently in cases where higher forecast probabilities were issued.
Such questions can be explored using diagnostics such as conditional verification of the ensemble mean, analyses of the spread-error relationship, and reliability diagrams for forecast probabilities.

In these conditional diagnostics, reliability is typically assessed by comparing the conditional relationship between an ensemble-derived predictor and the corresponding observed quantity to its ideal one-to-one (diagonal) relationship \parencite[][]{https://doi.org/10.2307/2346866}.
Equivalently, reliability can be quantified through the slope of a linear fit of this relationship, for example the slope of the mean observed anomaly conditioned on the ensemble mean, the slope of squared error conditioned on ensemble spread, or the slope of observed event frequency conditioned on forecast probability.
A slope equal to its ideal value is commonly interpreted as evidence of perfect reliability \parencite{brockerConceptExchangeabilityEnsemble2011}.
Assessing deviations from the ideal relationship is considered important both for model verification and development \parencite[e.g.,][]{haidenEvaluationECMWFForecasts2025} and for users, as probabilistic reliability underpins forecast-based decision-making and provides a quantitative basis for assessing the overall ``goodness'' and trustworthiness of ensemble prediction systems \parencite{palmerSeamlessPredictionCalibration2008, 10.1098/rsif.2013.1162}.

However, probability-based reliability diagrams are known to suffer not only from sampling uncertainty due to a finite number of forecast cases \parencite{brockerIncreasingReliabilityReliability2007}, but also from systematic finite-ensemble effects \parencite{Richardson2001, manzanasReliabilityGlobalSeasonal2022}.
Specifically, \textcite{Richardson2001} showed that this bias manifests as a clockwise tilt in reliability diagrams, that is, a flattening of the slope relating observed event frequency to forecast probability.
Recently, \textcite{ruppSpreadversuserrorFrameworkReliably2025} demonstrated that an analogous attenuation also affects the spread-error relationship.
Awareness of this systematic attenuation bias is therefore crucial, as failing to account for it may lead to the erroneous conclusion that a forecast system is not perfectly reliable, even in the absence of any genuine reliability deficiencies.

In this study, we build on the work by \textcite{ruppSpreadversuserrorFrameworkReliably2025} and provide a unified and generalized theoretical framework of finite-ensemble slope attenuation across conditional reliability diagnostics for ensemble mean, spread, and probabilities.
We then test this theory using synthetic forecast experiments in which ensemble members and observations are generated from prescribed population distributions, such that the forecast system is perfectly reliable by construction and free of model error.
In addition, the synthetic framework allows us to generate (almost) arbitrarily large numbers of forecast-observation pairs, thereby eliminating sampling uncertainty associated with a finite number of forecast cases and enabling a clean isolation of finite-ensemble effects.

We demonstrate that systematic slope attenuation arises in all three diagnostics -- conditional verification of the ensemble mean, the spread-error relationship, and reliability diagrams for forecast probabilities.
We explain the origin of these effects and relate them to the more general statistical phenomenon of \textit{slope attenuation bias} \parencite[also called \textit{regression dilution bias;}][]{snedecorCochranStatisticalMethods1967,frostCorrectingRegressionDilution2000}.
Furthermore, we show that the attenuated slopes can be computed in practice without reference to observational data, which enables the benchmark for forecast evaluation to be adjusted for finite-ensemble effects.
Finally, we apply the derived correction methods to ensemble forecasts from the sub-seasonal prediction system of the European Centre for Medium-Range Weather Forecasts (ECMWF), illustrating the practical implications of our findings in the context of the spread-error relationship and tercile-based reliability diagrams.
However, the framework and correction methods are general and can be applied to any ensemble forecasting system, at any lead time.

The remainder of this paper is structured as follows.
In Section~\ref{sec:notation}, we introduce the notation used throughout the paper.
Section~\ref{sec:synthetic_data_experiments} demonstrates the finite-ensemble slope attenuation effects using synthetic data experiments, explains the underlying mechanism, and derives ideal slope corrections.
Section~\ref{sec:ecmwf_application} applies the derived correction methods to ECMWF sub-seasonal forecasts, focusing on the spread-error relationship and tercile-based reliability diagrams.
Finally, Section~\ref{sec:discussion} discusses the implications of our findings, and Section~\ref{sec:conclusions} summarizes our conclusions.

\section{Notation}
\label{sec:notation}

\subsection{Ensemble-derived and observed quantities}

Let cases be indexed by $j = 1, \ldots, N$, for example different forecast start dates.
For case $j$, let $y_j$ denote the verifying observation and let
$\{ x_{1,j}, \ldots, x_{m,j} \}$ denote an ensemble forecast of size $m$.

The ensemble mean for case $j$ is defined as
\[
\bar{x}_j
=
\frac{1}{m}
\sum_{i=1}^{m}
x_{i,j},
\]
and the ensemble variance as
\[
s_j^{2}
=
\frac{1}{m-1}
\sum_{i=1}^{m}
\left( x_{i,j} - \bar{x}_j \right)^{2}.
\]
The squared error of the ensemble mean for case $j$ is given by
\[
e_j^2
=
\left( y_j - \bar{x}_j \right)^2,
\]
which serves as the verifying quantity in spread--error relationships.
Unless indicated otherwise, a bias-corrected version of $e_j^2$ is used, by multiplying with a factor of $\frac{m}{m+1}$, which is an unbiased estimator of the squared error that would be expected in the limit of infinite ensemble size \parencite{robertsUnbiasedCalculationEvaluation2025}:
\[
\tilde{e}_j^2 = \frac{m}{m+1} e_j^2.
\]

For a given event $A$, the ensemble-based probability for case $j$ is
\[
p_j(A)
=
\frac{1}{m}
\sum_{i=1}^{m}
\mathbb{I}\!\left( x_{i,j} \in A \right),
\]
where $\mathbb{I}(\cdot)$ denotes the indicator function.
For example, exceedance probabilities correspond to events of the form
$A = [z,\infty)$, where $z$ denotes a fixed threshold, while tercile probabilities
correspond to intervals defined by climatological quantiles, such as
$A = (q_{1/3}, q_{2/3}]$.

The corresponding observed event indicator for case $j$ is defined as
\[
o_j(A)
=
\mathbb{I}\!\left( y_j \in A \right),
\]
and serves as the verifying quantity in probability-based reliability diagrams.

To describe statistics across cases, for any sequence $a_j$ (such as $\bar{x}_j$, $s_j^2$, or $p_j$), we define the empirical expectation and variance over the case index $j$ as
\[
\mathbb{E}_j[ a_j ]
=
\frac{1}{N}
\sum_{j=1}^{N}
a_j,
\qquad
\mathrm{Var}_j( a_j )
=
\mathbb{E}_j\!\left[
\left( a_j - \mathbb{E}_j[a_j] \right)^2
\right].
\]

\subsection{Unobservable population parameters}

For each case $j$, ensemble members are assumed to be independent draws from a
case-specific forecast distribution, denoted by $F_j$, corresponding to the limit
of infinite ensemble size.
This distribution is not directly observable, but each ensemble member
$x_{i,j}$ is assumed to be an independent realization from $F_j$.
Similarly, the verifying observation $y_j$ is also assumed to be an independent realization from $F_j$, making the forecast system perfectly reliable by construction.

The distribution $F_j$ has population mean $\mu_j$,
population variance $\sigma_j^2$, and population probability $\pi_j(A)$
for a given event $A$.

\subsection{Conditional evaluation}

Throughout this paper, expectations and variances are used in two senses: to describe behaviour within a single forecast case, and to describe behaviour across the collection of forecast cases.

Expressions of the form
\[
\mathbb{E}[\cdot \mid j]
\]
refer to expectations under the case-specific forecast distribution $F_j$ for a fixed forecast case $j$. This expectation corresponds to repeated hypothetical realizations drawn from the same forecast distribution.
For example, for a fixed case $j$, the quantity
\[
\mathbb{E}[\bar{x}_j \mid j]
\]
denotes the expectation of the ensemble mean under the forecast distribution $F_j$, that is, the value the ensemble mean would converge to with infinite ensemble members.
Under the assumption of reliability (i.e., ensemble members $x_{i,j}$ are independent draws from $F_j$), this expectation equals the population mean $\mu_j$. This expectation is conceptual, as each forecast case is observed only once.

By contrast, expressions of the form
\[
\mathbb{E}_j[\cdot]
\]
denote averages taken over the index $j$, that is, across the collection of forecast cases in the dataset.
For example,
\[
\mathbb{E}_j[\bar{x}_j]
\]
denotes the average of the ensemble mean across different cases.

Conditional evaluation over subsets of cases is written as
\[
\mathbb{E}_j[\cdot \mid \mathcal{C}],
\]
where $\mathcal{C}$ denotes a condition imposed on observable or unobservable quantities. For example,
\[\mathbb{E}_j[\bar{x}_j \mid \bar{x}_j > 0]\]
denotes the average ensemble mean over all cases satisfying the condition $\bar{x}_j > 0$.

\section{Finite-ensemble slope attenuation: theory and synthetic data experiments}
\label{sec:synthetic_data_experiments}

To isolate and quantify the effect of finite ensemble size on conditional reliability diagnostics, we generate synthetic ensemble forecasts and observations that are perfectly reliable by construction, allowing any deviations from ideal diagnostic behaviour to be attributed solely to finite-ensemble sampling effects.
Similar approaches have been used in previous studies to investigate forecast verification issues \parencite[e.g.,][]{kruizingaKokEvaluationECMWF1988,grimitMeasuringEnsembleSpread2007,hopsonAssessingEnsembleSpread2014}.
The theoretical results derived below are not specific to this synthetic construction, but apply more generally to conditional reliability diagnostics in finite ensembles; the synthetic setup merely provides a controlled environment in which the mechanisms can be isolated and quantified.

\subsection{Experimental design of synthetic data experiments}
\label{sec:synthetic_data_experiments_experimental_design}

Synthetic ensemble forecasts and observations are generated as follows.
For each case $j$, a population distribution $F_j$ is defined as a normal distribution with mean $\mu_j$ and variance $\sigma_j^2$.
The population parameters $\mu_j$ and $\sigma_j^2$ are randomly drawn for each case: $\mu_j$ is drawn from a normal distribution with mean 0 and variance $\tau^2$, and $\sigma_j^2$ is drawn from a scaled chi-squared distribution with $\mathrm{df}$ degrees of freedom.

For each case $j$, given the population distribution $F_j$, an ensemble forecast of size $m$ is generated by drawing $m$ independent samples from $F_j$.
The verifying observation $y_j$ is also independently drawn from the same distribution $F_j$.
The (empirical) ensemble mean $\bar{x}_j$, ensemble variance $s_j^2$, and ensemble-based probability of exceedance $p_j$ are computed from the generated ensemble members.

The case-to-case variability of the population mean, $\mu_j$, is controlled by the parameter $\tau$, which we consider from the set $\{0.05, 0.15, 0.5\}$, and label as \emph{small}, \emph{medium}, and \emph{large}, respectively.
The case-to-case variability of the population variance, $\sigma_j^2$, is controlled by the degrees of freedom parameter, $\mathrm{df}$, which we consider from the set $\{150, 30, 7, 3\}$, and label as \emph{small}, \emph{medium}, \emph{large} and \emph{x-large}, respectively.

\[
\underbrace{
x_{i,j} \mid \mu_j,\sigma_j^2 \sim \mathcal{N}(\mu_j,\sigma_j^2),
\quad i=1,\ldots,m
}_{\text{ensemble members}}
\qquad
\underbrace{
y_j \mid \mu_j,\sigma_j^2 \sim \mathcal{N}(\mu_j,\sigma_j^2)
}_{\text{verifying observation}}
\]

\[
\underbrace{
\mu_j \sim \mathcal{N}(0,\tau^2)
}_{\text{population mean variability}}
\qquad
\underbrace{
\sigma_j^2 \sim \chi^2_{\mathrm{df}}/\mathrm{df}
}_{\text{population variance variability}}
\]

All distributions are illustrated in supplementary Fig.~S1.

Our results are not sensitive to the specific choice of population distribution.
In supplementary Fig.~S5 we show that our findings hold also when ensemble members and observations are drawn from a skew-normal distribution.

\subsection{Demonstrating slope attenuation in conditional verification}

We demonstrate the effect of slope attenuation in conditional verification of ensemble mean, spread, and probabilities using the synthetic data experiments.
The first set of experiments is based on $n=200\,000$ cases, with the underlying population distribution exhibiting medium variability in the mean ($\tau=0.15$) and medium variability in the variance ($\mathrm{df}=30$), while the ensemble size is varied from $m=11$ to $m=250$.

\begin{figure}[ht]
    \centering
    \includegraphics[width=\textwidth]{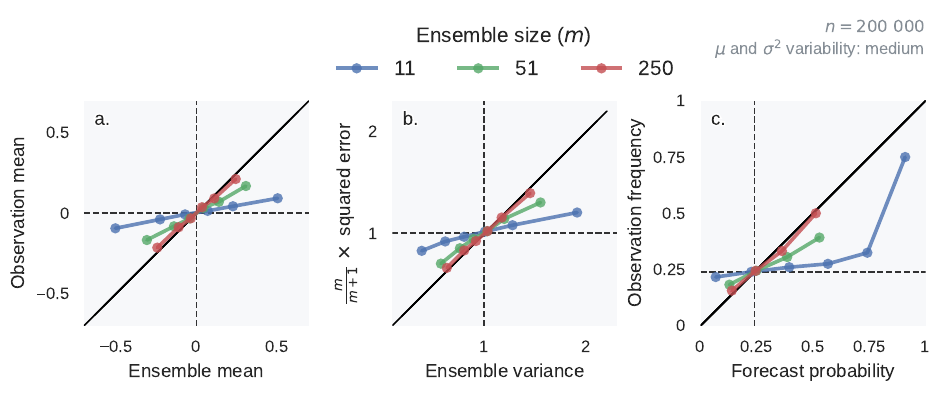}
    \caption{Illustration of slope attenuation in conditional verification due to finite ensemble size. (a) Conditional verification of ensemble mean: observed mean ($\mathbb{E}[y_j|\bar{x}_j]$) versus ensemble mean ($\bar{x}_j$) for different ensemble sizes $m$. (b) Spread-error relationship: unbiased squared error ($\frac{m}{m+1}  e_j^2$) versus ensemble variance ($s_j^2$) for different ensemble sizes $m$. (c) Reliability diagram: observed frequency ($\mathbb{E}[o_j|p_j]$) versus forecast probability ($p_j$) for different ensemble sizes $m$.
    In all panels, the ideal one-to-one relationship is indicated by the solid black diagonal line.
    Horizontal and vertical dashed lines indicate climatological means.}
    \label{fig:overview_mean_spread_p_different_m}
\end{figure}

In case of conditioning on the ensemble mean, all ensemble forecasts are stratified into six quantiles based on their ensemble mean values $\bar{x}_j$.
Note that the number of bins can be chosen more systematically depending on the number of cases and ensemble members \parencite[see, for example,][]{dimitriadisStableReliabilityDiagrams2021}. However, the exact choice is not crucial here, as the binned conditional means serve only as a visual aid to illustrate the underlying conditional relationship, while all analytical expressions derived in the following sections are based on an ordinary least-squares linear regression using the full, unbinned data.
For each quantile, the mean of the verifying observations $y_j$ is computed. The relationship between the ensemble mean quantiles and the corresponding observed means is then examined. This is repeated for different ensemble sizes $m$ to illustrate the effect of finite ensemble size on the conditional relationship.
Fig.~\ref{fig:overview_mean_spread_p_different_m}a shows that for an ensemble size of $m=250$, the conditional relationship between the ensemble mean and the observed mean closely follows the ideal one-to-one relationship expected for a perfectly reliable forecast system. However, as the ensemble size decreases to $m=51$ or even $m=11$, the slope of the observed mean versus ensemble mean relationship becomes significantly attenuated, deviating from the 1-to-1 line. This demonstrates that finite ensemble size leads to a systematic flattening of the conditional relationship, even though the underlying forecast system is perfectly reliable by construction.

Similarly, Fig.~\ref{fig:overview_mean_spread_p_different_m}b illustrates the effect of finite ensemble size on the spread-error relationship. Here, the ensemble forecasts are stratified into six quantiles based on their ensemble spread values $s_j^2$. For each quantile, the mean squared error between the ensemble mean and the verifying observations is computed and multiplied by a factor of $\frac{m}{m+1}$, which is an unbiased estimator of the squared error in the limit of infinite ensemble size \parencite{Leutbecher2008}.
The relationship between ensemble spread quantiles and the corresponding squared errors is examined for different ensemble sizes $m$. Again, we observe that as the ensemble size decreases, the slope of the squared error versus ensemble spread relationship becomes increasingly attenuated.
Importantly, this behaviour does not indicate systematic under-dispersion or over-dispersion of the ensemble.
Rather, it reflects a regression-to-the-mean effect \parencite{stiglerRegressionMeanHistorically1997} induced by conditioning on a noisy estimate of the population spread: cases with large ensemble spread tend to exhibit squared errors that are not as large as the spread would suggest, while cases with small ensemble spread tend to exhibit squared errors that are not as small as the spread would suggest.
As a result, finite ensemble size leads to a flattening of the conditional spread-error relationship.

Finally, Fig.~\ref{fig:overview_mean_spread_p_different_m}c shows the impact of finite
ensemble size on reliability diagrams for forecast probabilities \parencite[as described in ][]{Wilks2007}.
The ensemble forecasts are stratified into six probability bins based on their
ensemble-based probabilities $p_j(A)$ for a given event
$A = [z,\infty)$, with $z = 0.7$ denoting a fixed threshold.
This choice is arbitrary and corresponds to a climatological event probability of
approximately 25\%.
For each bin, the observed frequency of exceedance is computed.
The relationship between forecast probabilities and observed frequencies is examined
for different ensemble sizes $m$.
While the mean forecast probability remains unbiased across ensemble sizes and equals
the climatological event probability (approximately 25\%), finite ensemble size affects
the conditional relationship.
In particular, for small ensembles, cases with large forecast probabilities tend to be
associated with observed event frequencies that are not as large as indicated by the
forecast, while cases with small forecast probabilities tend to be associated with
observed frequencies that are not as small.
As a result, decreasing ensemble size leads to a pronounced attenuation of the slope in
the reliability diagram, demonstrating that finite ensemble size systematically flattens
the conditional relationship between forecast probabilities and observed frequencies.

It is important to note that in all three cases, the slope attenuation arises from the conditioning on ensemble-derived statistics, which are noisy estimates of the true population parameters due to the finite ensemble size. If conditioning were performed on the true population parameters (i.e., $\mu_j$, $\sigma_j^2$, and $\pi_j$), no slope attenuation would occur (see supplementary Fig.~S2).

\subsection{Explaining slope attenuation in conditional verification}

The slope attenuation observed in the conditional reliability diagnostics is an instance of the classical slope attenuation (or regression dilution) bias, which arises when a noisy estimate of an underlying quantity is used as the predictor variable \parencite{Fuller1987MeasurementError,carroll2006measurement}. In the context of ensemble forecasting, this noise originates from finite ensemble sizes.

To make this connection explicit, we represent ensemble-derived estimates as noisy observations of the true population parameters:
\[
\bar{x}_j = \mu_j + \epsilon_{\mu,j},
\qquad
s_j^2 = \sigma_j^2 + \epsilon_{\sigma,j},
\qquad
p_j = \pi_j + \epsilon_{\pi,j},
\]
where $\epsilon_{\mu,j}$, $\epsilon_{\sigma,j}$, and $\epsilon_{\pi,j}$ represent the noise due to finite ensemble size.
Remember that the population parameters $\mu_j$, $\sigma_j^2$, and $\pi_j$ are known only in our synthetic experiments, but are not observable in real-world scenarios.

Importantly, when the expectation is taken over all cases without conditioning, the noise terms have zero mean:
\[\mathbb{E}_j[\epsilon_{\mu,j}] = 0,
\qquad
\mathbb{E}_j[\epsilon_{\sigma,j}] = 0,
\qquad
\mathbb{E}_j[\epsilon_{\pi,j}] = 0.\]
These expressions imply that the forecast system is unbiased and well-calibrated (neither over- nor under-dispersive) on average across all cases.

However, the expectation of the noise terms no longer remains generally zero when conditioning on particular subsets of cases based on the noisy estimates:
\[\mathbb{E}[\epsilon_{\mu,j} \mid \bar{x}_j] \ne 0,
\qquad
\mathbb{E}[\epsilon_{\sigma,j} \mid s_j^2] \ne 0,
\qquad
\mathbb{E}[\epsilon_{\pi,j} \mid p_j] \ne 0.\]
This is because the conditioning effectively selects cases not only based on the true population parameters but also on the noise, leading to biased estimates of the population parameters in the conditioned subset.

To illustrate the mechanism, consider a centred setting in which the distribution
of population means $\mu_j$ has median zero, and partition the cases into two
quantiles based on the sign of the ensemble mean $\bar{x}_j$.
Because $\bar{x}_j = \mu_j + \epsilon_{\mu,j}$, this partition does not coincide with a partition
based on $\mu_j$.
In particular, some cases with $\mu_j > 0$ are assigned to the negative subset due
to negative sampling noise, while some cases with $\mu_j < 0$ are assigned to the
positive subset due to positive sampling noise.
As a result, the average ensemble mean in each subset is pulled toward zero, leading to an attenuation of the conditional slope.

An analogous argument applies to the ensemble spread and ensemble-based probabilities.
When conditioning on the ensemble spread $s_j^2$, cases with positive noise $\epsilon_{\sigma,j}$ are more likely to be included in the high-spread subset, while cases with negative noise are more likely to be included in the low-spread subset.
When conditioning on the ensemble-based probability $p_j$, cases with positive noise $\epsilon_{\pi,j}$ are more likely to be included in the high-probability subset, while cases with negative noise are more likely to be included in the low-probability subset.
This selection bias leads to an attenuation of the conditional slopes in all three cases.

\subsection{Ideal slope correction using population parameters}

Fig.~\ref{fig:overview_mean_spread_p_different_m} showed that conditioning on
ensemble-derived statistics leads to slope attenuation in conditional verification
of the ensemble mean, ensemble spread, and ensemble-based probabilities.
In practice, one would like to correct for this effect in order to obtain unbiased
estimates of the underlying conditional relationships.
The term ``correction'' refers to adjusting the benchmark against which the observed conditional relationship is compared, rather than modifying the data itself.
This benchmark equals one only in the limit of infinite ensemble size, while for finite ensembles it is expected to be attenuated.
In the following, we derive an analytical expression for the slope attenuation as a
function of the population parameters and the ensemble size.
We begin with the case of conditioning on the ensemble mean; the derivations for
conditioning on ensemble spread and probabilities follow analogously.

If the population parameters were known, conditional verification would recover a unit slope. Specifically, under the reliability assumption (i.e., ensemble members and observations are independent realizations from the population distribution $F_j$), the conditional expectation of the observation given the population mean satisfies $\mathbb{E}[y_j \mid \mu_j] = \mu_j$, so that the slope of a linear regression of $y_j$ on $\mu_j$ is equal to one.
Note that this regression is based on the full, unbinned data rather than on binned conditional means; the latter typically provide a close approximation but may differ slightly when the number of bins and cases is finite.
In practice, however, conditioning is performed on the finite-ensemble estimate
$\bar{x}_j$, which is related to the population mean by
$\bar{x}_j = \mu_j + \epsilon_{\mu,j}$, where $\epsilon_{\mu,j}$ denotes
finite-ensemble sampling noise.

Conditioning on $\bar{x}_j$ therefore mixes cases with different underlying
population means, leading to an attenuation of the conditional slope.
The resulting slope of the linear regression of $y_j$ on $\bar{x}_j$ can be written as
\begin{align}
\mathrm{slope}\!\left(\mathbb{E}[y_j \mid \bar{x}_j],\, \bar{x}_j\right)
&=
\frac{\mathrm{Cov}_j(\mathbb{E}[y_j \mid \bar{x}_j],\, \bar{x}_j)}
     {\mathrm{Var}_j(\bar{x}_j)} \notag \\[0.6ex]
&=
\frac{\mathrm{Cov}_j(\mathbb{E}[\mu_j \mid \bar{x}_j],\, \bar{x}_j)}
     {\mathrm{Var}_j(\bar{x}_j)} \notag \\[0.6ex]
&=
\frac{\mathrm{Cov}_j(\mu_j,\, \bar{x}_j)}
     {\mathrm{Var}_j(\bar{x}_j)} \notag \\[0.6ex]
&=
\frac{\mathrm{Var}_j(\mu_j)}
     {\mathrm{Var}_j(\mu_j) + \mathrm{Var}_j(\epsilon_{\mu,j})} \notag \\[0.6ex]
&=
1 - \frac{\mathrm{Var}_j(\epsilon_{\mu,j})}{\mathrm{Var}_j(\bar{x}_j)} .
\label{eq:slope_attenuation_mean}
\end{align}
Here we used that, under the reliability assumption, the observation can be written as
$y_j = \mu_j + \eta_j$, where $\eta_j$ denotes observational variability with
zero mean conditional on $\mu_j$.
Since $y_j$ depends on the ensemble mean $\bar{x}_j$ only through the underlying
population mean $\mu_j$, averaging $y_j$ over cases with the same $\bar{x}_j$
removes the observational variability and yields
$\mathbb{E}[y_j \mid \bar{x}_j] = \mathbb{E}[\mu_j \mid \bar{x}_j]$.
The covariance simplification follows from the identity
$\mathrm{Cov}(\mathbb{E}[U \mid V], V) = \mathrm{Cov}(U,V)$, as a special case of the law of total covariance.
Finally, writing $\bar{x}_j = \mu_j + \epsilon_{\mu,j}$, the variance decomposition
follows because the ensemble sampling noise $\epsilon_{\mu,j}$ arises solely from
finite sampling of the case-specific forecast distribution and is therefore
uncorrelated with the underlying population mean $\mu_j$ across cases.

This derivation is not specific to the ensemble mean, but applies to any noisy
estimate of a population quantity.
In particular, the ensemble spread estimate satisfies
$s_j^2 = \sigma_j^2 + \epsilon_{\sigma,j}$, and the ensemble-based probability
estimate satisfies $p_j(A) = \pi_j(A) + \epsilon_{\pi,j}$.
Consequently, conditional slopes in the spread--error relationship and in reliability
diagrams for probabilities are attenuated in an analogous manner:
\begin{equation}
\mathrm{slope}\!\left(\mathbb{E}[\tilde{e}_j^2 \mid s_j^2],\, s_j^2\right)
=
\frac{\mathrm{Var}_j(\sigma_j^2)}
     {\mathrm{Var}_j(\sigma_j^2) + \mathrm{Var}_j(\epsilon_{\sigma,j})}
=
1 - \frac{\mathrm{Var}_j(\epsilon_{\sigma,j})}{\mathrm{Var}_j(s_j^2)},
\label{eq:slope_attenuation_spread}
\end{equation}
and
\begin{equation}
\mathrm{slope}\!\left(\mathbb{E}[o_j(A) \mid p_j(A)],\, p_j(A)\right)
=
\frac{\mathrm{Var}_j(\pi_j(A))}
     {\mathrm{Var}_j(\pi_j(A)) + \mathrm{Var}_j(\epsilon_{\pi,j})}
=
1 - \frac{\mathrm{Var}_j(\epsilon_{\pi,j})}{\mathrm{Var}_j(p_j(A))}.
\label{eq:slope_attenuation_probability}
\end{equation}

In all cases, slope attenuation is stronger when the variance of the finite-ensemble
noise is large relative to the variability of the corresponding population quantity.
For a given variability of the population parameters, the noise variances decrease
with increasing ensemble size $m$, leading to weaker slope attenuation for larger
ensembles.
Note that the derived expressions correspond to the slope of the least-squares
linear regression line computed from all forecast cases without binning; slopes estimated from binned conditional means or frequencies can differ slightly when the number of bins and the number of cases are finite.

\begin{figure}[ht]
    \centering
    \includegraphics[width=\textwidth]{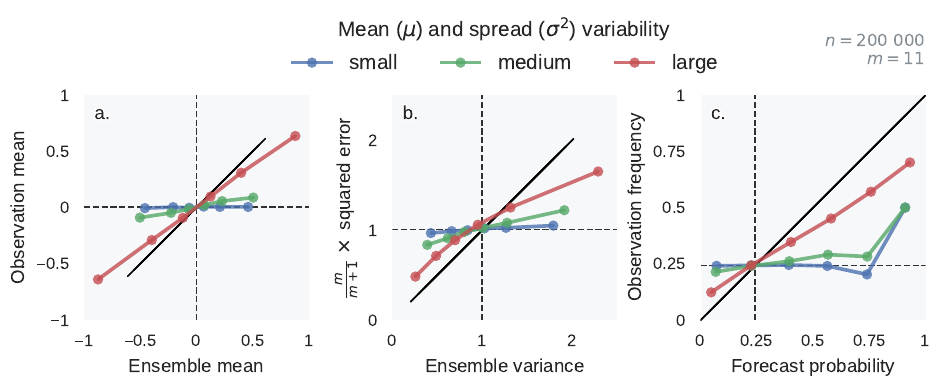}
    \caption{Following Fig.~\ref{fig:overview_mean_spread_p_different_m}, but varying the variability of the population parameters instead of the ensemble size, using a fixed ensemble size of $m=11$.
    The population mean, $\mu_j$, for case $j$ is drawn from a normal distribution with mean 0 and variance $\tau^2$, where $\tau$ equals 0.05 (small), 0.15 (medium), and 0.5 (large). 
    Likewise, the population variance, $\sigma_j^2$, for case $j$ is drawn from a chi-squared distribution with $\mathrm{df}$ degrees of freedom, where $\mathrm{df}$ equals 150 (small), 30 (medium), and 7 (large).
    }
    \label{fig:overview_mean_spread_p_different_s}
\end{figure}

Fig.~\ref{fig:overview_mean_spread_p_different_s} illustrates the effect of varying the variability of the population parameters instead of the ensemble size, using a fixed ensemble size of $m=11$.
As expected from Eqs.~\eqref{eq:slope_attenuation_mean}-\eqref{eq:slope_attenuation_probability}, larger variability of the population parameters leads to less slope attenuation, while smaller variability leads to stronger slope attenuation.
In the limiting case where the population parameters are constant across cases (i.e., no case-to-case variability; not shown), even an infinite ensemble would collapse to a single value on the conditioning axis, yielding no conditional relationship.
In finite ensembles, residual variability along the conditioning axis arises solely from sampling noise and is therefore uncorrelated with variations in observations or errors, resulting in a vanishing slope.

We now evaluate the analytical expressions in Eqs.~\eqref{eq:slope_attenuation_mean}-\eqref{eq:slope_attenuation_probability} using synthetic data experiments and
compare them to empirical slopes obtained from least-squares fits between the
ensemble-derived estimates and the corresponding verifying quantities.
In these synthetic experiments, the finite-ensemble noise terms are known exactly,
as they can be computed directly as the difference between the ensemble-derived
statistics and the prescribed population quantities.
Figure~\ref{fig:slope_estimators} shows that for all three verification settings -- ensemble mean, ensemble spread, and ensemble-based
probabilities -- the analytically derived slopes closely match the empirical slopes
obtained from conditional verification.
In particular, the dependence of the slope on ensemble size $m$, including the
systematic decrease of the slope for smaller ensembles, is well captured by the
analytical expressions.

\subsection{Estimating slope corrections from finite ensembles}
\label{subsec:estimating_slope_corrections}

Equations~\eqref{eq:slope_attenuation_mean}-\eqref{eq:slope_attenuation_probability}
provide analytical expressions for the attenuation of conditional slopes due to
finite-ensemble sampling noise.
These expressions quantify the ratio between the empirically observed slope
obtained when conditioning on ensemble-derived statistics and the corresponding
population-level slope, which equals one under the reliability assumptions.
As such, they can be interpreted as ideal slope correction factors, assuming
knowledge of the population parameters and noise variances.
In practice, neither the population parameters nor the noise variances are known, making Eqs.~\eqref{eq:slope_attenuation_mean}-\eqref{eq:slope_attenuation_probability} conceptually illuminating but not directly applicable in an operational forecasting context.
However, we will now propose estimators for the noise variances based on the ensemble
data, allowing us to estimate the slope correction factors from ensemble data alone, without knowledge of the population parameters or verifying observations.

\subsubsection{Ensemble mean versus observed mean}

We begin with the ensemble mean.
For a single case $j$, we use the ensemble mean as an unbiased but noisy estimator
of the population mean.
The magnitude of this noise is directly related to the ensemble variance for that
case.
Specifically, the variance of the noise in the ensemble mean for case $j$ is given
by the squared standard error of the mean:
\[
\mathrm{Var}(\epsilon_{\mu,j} \mid j)
=
\mathrm{Var}(\bar{x}_j \mid j)
=
\frac{\sigma_j^2}{m}.
\]

While the population variance $\sigma_j^2$ is unknown, it can be estimated by the ensemble variance $s_j^2$, which is an unbiased estimator of $\sigma_j^2$, so that
\[\widehat{\mathrm{Var}(\epsilon_{\mu,j} \mid j)}
=
\frac{s_j^2}{m}.\]
Here and in the following, a hat indicates that the corresponding expression is a sample-based estimator of the underlying population quantity.
Supplementary Fig.~S3a validates this estimator using the synthetic data experiments, showing that it closely matches the true noise variance.
Averaging over cases yields the average noise variance across the verification
sample,
\[
\mathrm{Var}_j(\epsilon_{\mu,j})
=
\mathbb{E}_j\!\left[\mathrm{Var}(\bar{x}_j \mid j)\right]
=
\mathbb{E}_j\left[ \frac{s_j^2}{m} \right].
\]

Using this estimate in Eq.~\eqref{eq:slope_attenuation_mean} yields the following
estimator of the conditional slope:
\begin{equation}
\widehat{\mathrm{slope}}\!\left(\mathbb{E}[y_j \mid \bar{x}_j],\, \bar{x}_j\right)
=
1 - \frac{\mathbb{E}_j\!\left[s_j^2/m\right]}{\mathrm{Var}_j(\bar{x}_j)}.
\label{eq:estimated-slope-mean}
\end{equation}
This estimator can be computed directly from the ensemble data without
knowledge of the population parameters or the verifying observations.

% \subsubsection{Ensemble spread versus squared error}

% We continue with the spread-error relationship. For a single case $j$, the variance of the noise in the ensemble variance estimate is given by
% \[
% \mathrm{Var}\!\left(\epsilon_{\sigma,j}\mid j\right)
% =
% \mathrm{Var}\!\left(s_j^2 \mid j\right)
% =
% \frac{(m-1)^2}{m^3}\left(\mu_{4,j}-\frac{m-3}{m-1}\sigma_j^{4}\right),
% \]
% where $\mu_4$ denotes the fourth central moment of the population distribution $F_j$ \parencite{raoLinearStatisticalInference1973}.
% Since the population quantities are unknown, we replace the expression
% \[
% \mu_{4,j}-\frac{m-3}{m-1}\sigma_j^4
% \]
% by its sample-based counterpart
% \[
% \hat{\mu}_{4,j}-\frac{m-3}{m-1}s_j^4,
% \]
% yielding a plug-in estimator of the noise variance, where
% \[
% \hat{\mu}_{4,j} = \frac{1}{m} \sum_{i=1}^{m} (x_{i,j} - \bar{x}_j)^4
% \]is the sample fourth central moment of the ensemble members for case $j$.

\subsubsection{Ensemble spread versus squared error}

We next consider the spread--error relationship.
For a fixed case \(j\), we require the conditional sampling variance of the ensemble variance, denoted by \(\mathrm{Var}(\epsilon_{\sigma,j}\mid j)\).
This quantifies the sampling noise in \(s_j^2\) that arises because only \(m\) ensemble members are available.
Equivalently, it is the variance that would be obtained by repeatedly drawing ensembles of size \(m\) from the same case-specific population distribution \(F_j\) and recomputing the sample variance each time.

Under the assumption that the ensemble members are independent draws from \(F_j\), the sampling variance of the unbiased sample variance is a standard result \parencite[see, e.g., p. 229 of][]{moodIntroductionTheoryStatistics1974}:
\[
\mathrm{Var}(\epsilon_{\sigma,j}\mid j)
=
\mathrm{Var}(s_j^2 \mid j)
=
\frac{1}{m}\left(\mu_{4,j}-\frac{m-3}{m-1}\sigma_j^{4}\right),
\]
where \(\mu_{4,j}\) denotes the fourth central moment of the case-specific population distribution \(F_j\).
Thus, unlike the sampling variance of the sample mean, which depends on the second central moment, the sampling variance of \(s_j^2\) depends on a fourth-order population moment.

If only ensemble data are available, the population quantities \(\sigma_j^4\) and \(\mu_{4,j}\) are unknown and must be estimated from the ensemble members.
A naive plug-in replacement is not, in general, exactly unbiased, because \((s_j^2)^2\) is not an unbiased estimator of \(\sigma_j^4\).
However, assuming that the ensemble members are independent draws from \(F_j\) and using standard \(k\)-statistics results \parencite[e.g., p. 189 of][]{kenneyMathematicsStatistics1947}, an unbiased estimator of \(\mathrm{Var}(s_j^2 \mid j)\) is given by
\[
\widehat{\mathrm{Var}(s_j^2\mid j)}
=
\frac{m}{(m-2)(m-3)}\,m_{4,j}
-
\frac{m^2-3}{m(m-2)(m-3)}\,(s_j^2)^2,
\qquad m \ge 4,
\]
where
\[
m_{4,j} = \frac{1}{m} \sum_{i=1}^m (x_{i,j} - \bar{x}_j)^4
\]
is the sample fourth central moment for case \(j\).
Supplementary Fig.~S3b validates this estimator using the synthetic data experiments, showing that it closely matches the true noise variance.

Substituting this estimator into Eq.~\eqref{eq:slope_attenuation_spread} yields an ensemble-based estimator of the expected attenuated slope in the spread--error relationship. This is the slope expected under perfect reliability at finite ensemble size:
\begin{equation}
\label{eq:estimated-slope-spread}
\widehat{\mathrm{slope}}\!\left(\mathbb{E}[\tilde e_j^2\mid s_j^2],\, s_j^2\right)
=
1-
\frac{
\mathbb E_j\!\left[\widehat{\mathrm{Var}(s_j^2\mid j)}\right]
}{
\mathrm{Var}_j(s_j^2)
}
=
1 - \frac{\mathbb{E}_j\left[\frac{m}{(m-2)(m-3)}\,m_{4,j} - \frac{m^2-3}{m(m-2)(m-3)}\,(s_j^2)^2\right]}{\mathrm{Var}_j(s_j^2)}.
\end{equation}
Hence, the expected slope attenuation can be estimated directly from ensemble moments up to order four, without using the verifying observations.

\subsubsection{Predicted event probabilities versus observed frequencies}
\label{subsubsec:estimated_slope_corrections_probabilities}

We now turn to the probability-based reliability evaluation. For a single case $j$, the sampling variance of the ensemble-based probability estimate follows from the binomial distribution \parencite[e.g.,][]{siegelChapter8Random2012}:
\[\mathrm{Var}(\epsilon_{\pi,j} \mid j) = \mathrm{Var}(p_j \mid j) = \frac{\pi_j(1 - \pi_j)}{m}.\]

The population probability $\pi_j$ is, in general, unknown, but the ensemble-based probability $p_j$ is an unbiased estimator of $\pi_j$.
However, the quantity $\pi_j(1-\pi_j)$ cannot be estimated directly by replacing $\pi_j$ with $p_j$, because $p_j^2$ is not an unbiased estimator of $\pi_j^2$.
Instead, it follows from the first two moments of the binomial distribution \parencite[e.g., Ch. 7.2 of][]{casellaBergerStatisticalInference2002} that an unbiased estimator of $\pi_j(1-\pi_j)$ is given by $\frac{m}{m-1}p_j(1-p_j)$, so that an unbiased estimator of the noise variance is
\[
\widehat{\mathrm{Var}(\epsilon_{\pi,j} \mid j)} = \frac{p_j(1 - p_j)}{m-1}.
\]
Supplementary Fig.~S3c validates this estimator using the synthetic data experiments, showing that it closely matches the true noise variance.

Averaging over all cases, we obtain an estimator for the noise variance across cases:
\[\mathrm{Var}_j(\epsilon_{\pi,j}) = \mathbb{E}_j\left[\mathrm{Var}(p_j \mid j)\right] = \mathbb{E}_j\left[\frac{p_j(1 - p_j)}{m-1}\right].\]
This is our estimator for the noise variance in the ensemble-based probability estimate.
This estimator can be used in equation~\eqref{eq:slope_attenuation_probability} to estimate the expected slope of the observed-frequency versus ensemble-based probability relationship:
\begin{equation}
    \label{eq:estimated-slope-probability}
    \widehat{\mathrm{slope}}(\mathbb{E}[o_j \mid p_j], p_j) = 1 - \frac{\mathbb{E}_j\left[\frac{p_j(1 - p_j)}{m-1}\right]}{\mathrm{Var}_j(p_j)}.
\end{equation}

A summary of the derived estimators for the slope corrections is provided in Table~\ref{tab:slope_corrections}.

\begin{table}[tbh]
    \centering
    \begin{tabular}{l l l l}
        \toprule
        & Population quantity & Ensemble-based estimate & $\mathrm{Var}(\epsilon_j \mid j)$ \\
        \midrule
        Mean & $\mu_j$ & $\bar{x}_j = \mu_j + \epsilon_{\mu,j}$ & $\frac{s_j^2}{m}$ \\
        Variance & $\sigma_j^2$ & $s_j^2 = \sigma_j^2 + \epsilon_{\sigma,j}$ & $\frac{m}{(m-2)(m-3)}\,m_{4,j} - \frac{m^2-3}{m(m-2)(m-3)}\,(s_j^2)^2$ \\
        Probability & $\pi_j$ & $p_j = \pi_j + \epsilon_{\pi,j}$ & $\frac{p_j(1 - p_j)}{m-1}$ \\
        \bottomrule
    \end{tabular}
    \caption{Estimators for the noise variances of ensemble-derived statistics.}
    \label{tab:slope_corrections}
\end{table}

\subsubsection{Validation of analytical slope expressions}

\begin{figure}[ht]
    \centering
    \includegraphics[width=\textwidth]{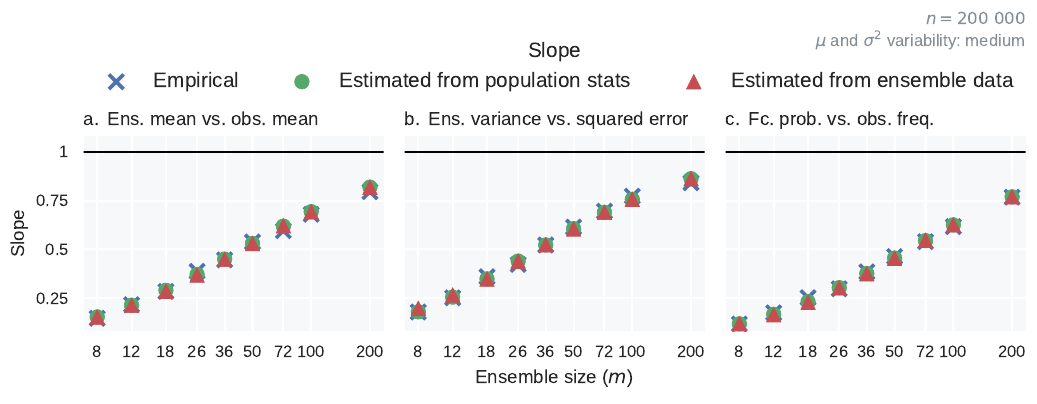}
    \caption{Comparison of empirical, ideal, and estimated slopes for varying ensemble sizes $m$. Panels (a), (b) and (c) are based on the same conditional diagnostics presented in Figs.~\ref{fig:overview_mean_spread_p_different_m}a, \ref{fig:overview_mean_spread_p_different_m}b, and \ref{fig:overview_mean_spread_p_different_m}c, respectively.
    Empirical slopes are computed by regressing observational statistics (observed mean, unbiased squared error, or event frequencies) on ensemble statistics (mean, variance, or probabilities).
    Slopes estimated from population statistics are computed using equations~\eqref{eq:slope_attenuation_mean}--\eqref{eq:slope_attenuation_probability} with true population parameters.
    Slopes estimated from ensemble data are computed using the derived estimators for noise variances following equations~\eqref{eq:estimated-slope-mean}--\eqref{eq:estimated-slope-probability}.
    Note the logarithmic scale on the horizontal axis.
    }
    \label{fig:slope_estimators}
\end{figure}

In Fig.~\ref{fig:slope_estimators}, we compare the slopes derived via three different methods: (i) empirical slopes computed directly from the observations, (ii) ideal slopes computed using the true population parameters (following Eqs.~\eqref{eq:slope_attenuation_mean}-\eqref{eq:slope_attenuation_probability}), and (iii) estimated slopes computed using only available ensemble data (following Eqs.~\eqref{eq:estimated-slope-mean}-\eqref{eq:estimated-slope-probability}).
Slopes are computed for varying ensemble sizes $m$, and for each relationship: observed mean versus ensemble mean, squared error versus ensemble spread, and observed frequency versus ensemble-based probability.
We find that the estimated slopes closely match the empirical slopes, demonstrating the effectiveness of the derived estimators for slope correction.

\subsection{Category-dependent slope attenuation in quantile-based reliability diagrams}

Reliability diagrams for tercile-based forecasts assess the conditional relationship
between ensemble-derived probabilities for each tercile category and the corresponding
observed event frequencies.
For each tercile, cases are stratified according to the forecast probability assigned
to that category, and the observed relative frequency of occurrence is evaluated within
each probability bin.
Under perfect reliability, one might expect the conditional relationship between forecast probabilities and observed frequencies to follow the diagonal with unit slope for all terciles. However, our results showed, consistent with \textcite{Richardson2001}, that finite-ensemble sampling noise leads to slope attenuation in reliability diagrams.

In the following, we highlight an additional implication of our results: different terciles can appear to exhibit different degrees of reliability, even though the underlying forecast system is perfectly reliable by construction.
Using two synthetic examples, we show that the centre tercile can exhibit markedly different degrees of slope attenuation than the outer terciles, which can give the impression of category-dependent reliability despite identical population-level reliability.

\begin{figure}[ht]
    \centering
    \includegraphics[width=.8 \textwidth]{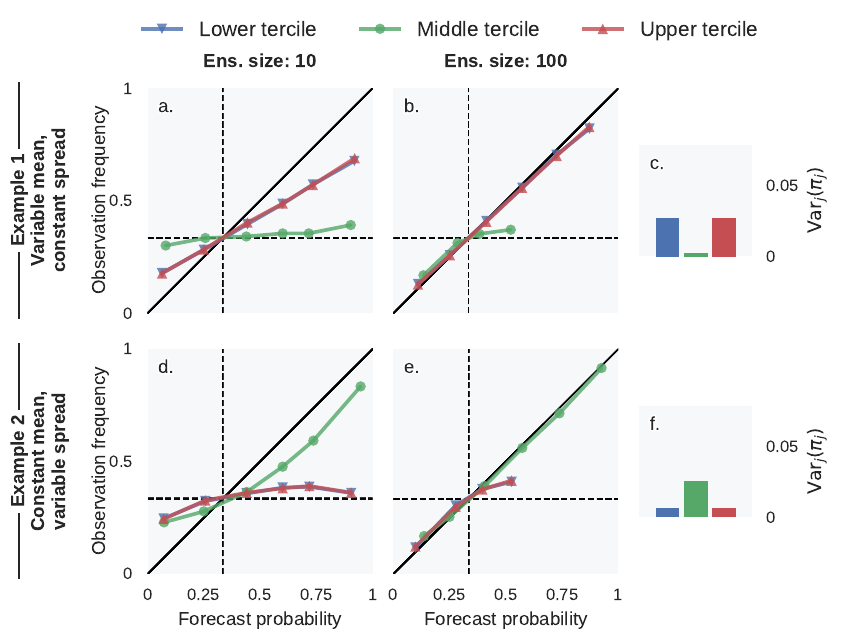}
    \caption{Illustration of category-dependent slope attenuation in tercile-based reliability diagrams for two idealized examples.
    (a--c) Example 1: population mean ($\mu_j$) varies across cases (large variability, that is, $\tau=0.5$) while the population variance ($\sigma^2_j$) does not vary across cases but is fixed at $\sigma^2_j=1$.
    (d--f) Example 2: population variance ($\sigma^2_j$) varies across cases (x-large variability, that is, $\mathrm{df}=3$) while the population mean ($\mu_j$) does  not vary across cases but is fixed at $\mu_j=0$.
    (a,d) Probability-based reliability diagrams for ensemble size $m=10$.
    (b,e) Same, but for ensemble size $m=100$.
    (c,f) Variance of population tercile probabilities across cases, $\mathrm{Var}_j(\pi_{k,j})$, for each tercile $k$ (lower, middle, upper).
    In all reliability diagrams, solid black diagonal lines show the 1-to-1 relationship. Dashed black lines indicate the climatological mean probability of 1/3. Results are based on $n=200\,000$ synthetic cases.}
    \label{fig:tercile_reliability_diagrams}
\end{figure}

Fig.~\ref{fig:tercile_reliability_diagrams} depicts these effects for two synthetic
setups with contrasting population variability characteristics.
In both examples, each case $j$ is associated with a Gaussian population distribution,
so that differences between terciles arise solely from variations in the first two
moments of the distribution.

In the first example, the population ensemble variance is fixed across cases, while the population mean varies.
In this setting, changes in the population mean predominantly affect the probabilities of the lower and upper terciles, whereas the probability of the central tercile remains comparatively stable.
As a result, the outer terciles exhibit larger case-to-case variability in their population probabilities and consequently show weaker slope attenuation, therefore appearing more reliable than the central tercile in the reliability diagram.

In the second example, the population mean is held fixed, while the population variance varies across cases.
Here, changes in scale primarily redistribute probability mass between the central and outer terciles.
The probability of the central tercile now exhibits the largest variability across cases, while the outer terciles remain closer to their climatological values.
Accordingly, the central tercile shows the weakest slope attenuation and appears most reliable, whereas the outer terciles exhibit stronger flattening of the conditional relationship.

These examples illustrate a general and potentially counter-intuitive consequence of finite-ensemble conditioning. Although terciles are defined such that their long-term mean probabilities are identical ($1/3$), the variability of their population probabilities across cases need not be.
Because slope attenuation is governed by the relative magnitude of sampling noise compared to the variability of the conditioning variable, categories whose population probabilities explore a wider range of values across cases are less affected by regression dilution and therefore appear more reliable.
Conversely, categories whose probabilities remain close to their climatological values exhibit stronger attenuation.

Consequently, differences in reliability slopes across terciles do not necessarily indicate genuine category- or flow-dependent deficiencies in the prediction system.
Instead, they arise from the interaction between finite-ensemble sampling noise and the distribution of population probability variability across categories.
Which tercile appears most reliable depends on the dominant mode of population variability, even when the underlying forecast system is perfectly reliable by construction.
Increasing the ensemble size reduces the overall magnitude of slope attenuation for all terciles, but does not eliminate the relative differences between categories (see Figs.~\ref{fig:tercile_reliability_diagrams}b,e).

\section{Application to ECMWF sub-seasonal forecasts}
\label{sec:ecmwf_application}

To illustrate the practical relevance of finite-ensemble slope attenuation in an operational forecasting context, we apply the proposed framework to ensemble forecasts from the European Centre for Medium-Range Weather Forecasts (ECMWF) sub-seasonal prediction system. We focus on weekly-mean 2-metre temperature anomaly forecasts at lead time week 4.

The analysis is based on ECMWF re-forecasts associated with real-time forecasts issued between January and April 2025.
For each real-time forecast date, the corresponding re-forecast set spans the period 2005-2024. Forecasts are initialized every two days (1 January, 3 January, ..., 29 April), reflecting the operational re-forecast schedule at the time of writing. The ensemble consists of 11 members, including one unperturbed control forecast and 10 perturbed members. In the following, only the 10 perturbed members are considered, ensuring that members are statistically exchangeable, consistent with the assumptions underlying the theoretical framework.
Anomalies are computed ``by member'', as described in \textcite{robertsUnbiasedCalculationEvaluation2025}.

\subsection{The spread-error relationship in ECMWF sub-seasonal forecasts}
\label{subsec:spread_error_ecmwf}

We begin by analysing the spread-error relationship in ECMWF sub-seasonal forecasts, which assesses the relationship between the ensemble variance and the unbiased squared error of the ensemble mean forecast. 
Slopes are computed for each month separately, then averaged over all four months to obtain robust estimates.
This ensures that variability associated with the seasonal cycle does not overly contribute to the diagnosed spread-error relationship.

Before analysing real-world forecast errors, we first adopt a perfect-model approach.
In this setting, one ensemble member is treated as the verifying observation, while the remaining members form the forecast ensemble. This procedure is repeated by iterating over all members, thereby generating a set of internally consistent forecast-verification pairs. The perfect-model framework is particularly useful here because it represents a system that is reliable at the population level by construction, while at the same time retaining the finite ensemble size of the operational re-forecast system (minus 1). As such, it allows finite-ensemble effects on the spread-error relationship to be isolated from additional sources of model error or observational uncertainty.

\begin{figure}[ht]
    \centering
    \includegraphics[width=\textwidth]{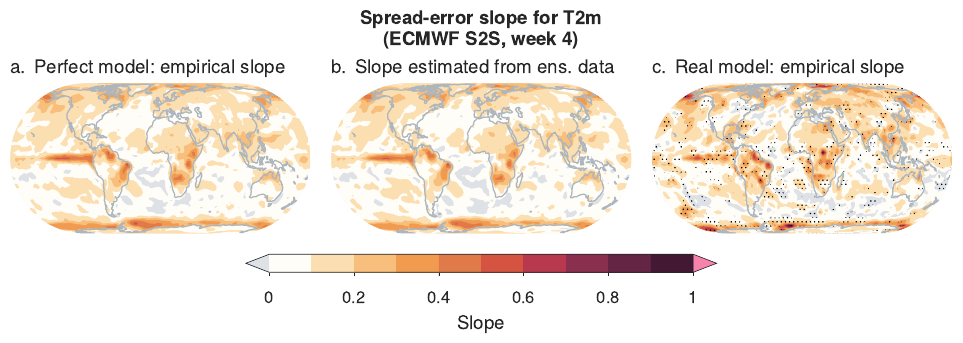}
    \caption{Spread-error slope for ECMWF sub-seasonal week-4 2-metre temperature forecasts (regressing unbiased squared error on ensemble variance) computed per calendar date, then averaged over forecasts initialized in January to April, from  2005-2024 (see text for details).
    (a) Empirical spread-error slope from perfect-model experiments, i.e., using one ensemble member as observation and the remaining members as forecast.
    (b) Estimated spread-error slope using the ensemble-based estimator (equation~\eqref{eq:estimated-slope-spread}), i.e., not using any information about observations and squared errors.
    (c) Empirical spread-error slope from real ECMWF forecasts verified against ERA5 reanalysis. Black stippling indicates grid points where the empirical slope is significantly different from the estimated slope at the 95\% confidence level based on resampling forecast start dates with replacement 200 times.
    All slopes are computed by iterating over all nine-member sub-ensembles, then averaging the results.
    }
    \label{fig:spread_error_slope_ecmwf}
\end{figure}

Figure~\ref{fig:spread_error_slope_ecmwf}a shows the empirically estimated spread-error slope obtained from the perfect-model experiments. Despite the absence of any true forecast error in a population sense, the diagnosed slopes deviate systematically from the ideal value of one, reflecting finite-ensemble slope attenuation. Figure~\ref{fig:spread_error_slope_ecmwf}b shows the corresponding slopes predicted by the ensemble-based estimator, following Eq.~\ref{eq:estimated-slope-spread} derived in section~\ref{subsec:estimating_slope_corrections}. The two fields are in near-perfect agreement, with a pattern correlation exceeding 0.99, demonstrating that the estimator accurately captures both the magnitude and spatial structure of slope attenuation arising purely from finite-ensemble effects.

Figure~\ref{fig:spread_error_slope_ecmwf}c shows the spread-error slope computed from ECMWF forecasts using ERA5 reanalysis as verification \parencite[see][for a similar analysis]{ruppSpreadversuserrorFrameworkReliably2025}. To maintain comparability with the perfect-model results, the analysis is again performed by iterating over all nine-member sub-ensembles. As expected, the resulting slopes are noisier and exhibit greater spatial variability, reflecting the presence of genuine forecast errors and additional sampling uncertainty due to only a single verifying observation per case. Nevertheless, the large-scale patterns remain broadly consistent with those obtained in the perfect-model framework.
Only about 15\% of the area exhibits slopes that are significantly different from the ensemble-based estimator at the 95\% confidence level, based on 200 bootstrap resamples of forecast start dates, indicating that the observed spread-error relationship is largely consistent with expectations from finite-ensemble effects alone.

Several regional features, common in Figs.~\ref{fig:spread_error_slope_ecmwf}a-c, are noteworthy. Enhanced slopes are found in regions associated with large-scale climate variability, such as the ENSO region, suggesting increased variability of the underlying population variance. More generally, spread-error slopes tend to be larger over land than over ocean, consistent with stronger case-to-case variability of near-surface temperature variability over continental regions. While such patterns may reflect genuine physical differences, their interpretation requires caution, as the diagnosed slopes represent a combination of true variability and finite-ensemble attenuation effects.

Overall, this application demonstrates that finite-ensemble slope attenuation is not merely a theoretical concern but has a tangible impact on spread-error diagnostics in operational sub-seasonal ensemble forecasts. The close agreement between perfect-model results and the ensemble-based estimator provides compelling evidence for the proposed correction framework. Moreover, the analysis highlights once again the importance of not interpreting an attenuated spread-error slope as evidence of model deficiency without carefully accounting for finite-ensemble effects.

\subsection{Tercile probability-based reliability diagrams for ECMWF sub-seasonal forecasts}
\label{subsec:reliability_diagrams_ecmwf}

In this section, we present tercile-based reliability diagrams for ECMWF sub-seasonal week-4 2-metre temperature weekly-mean anomaly forecasts, focusing on the middle and upper terciles.

The reliability diagrams are constructed as follows.
For each calendar date, the 33rd and 66th percentiles are computed across all cases (i.e., across re-forecast years) and ensemble members, defining tercile thresholds that vary by calendar date but are fixed across cases for a given date.
For each forecast case, ensemble-based tercile probabilities are then obtained as the fraction of ensemble members falling into the corresponding category.

\begin{figure}[ht]
    \centering
    \includegraphics[width=0.9\textwidth]{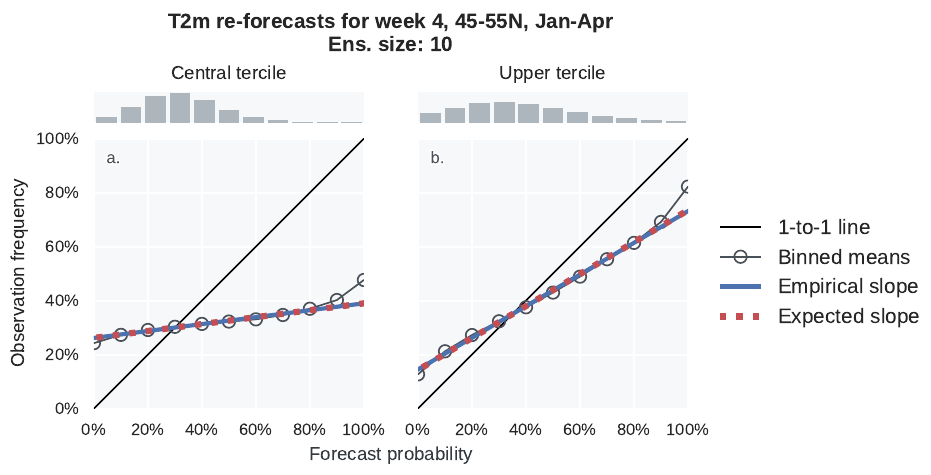}
    \caption{Tercile-based reliability diagrams for ECMWF sub-seasonal week-4 2-metre temperature weekly-mean anomaly forecasts, based on 10-member re-forecasts from 2005-2024 with forecasts initialized every two days from January to April (see text for details).
    (a) Middle tercile. (b) Upper tercile.
    Binned means are computed by stratifying cases into 11 forecast probability bins.
    Empirical slopes are computed by regressing observed frequencies on ensemble-based probabilities across all cases.
    Expected slopes are computed using the ensemble-based estimator derived in Section~\ref{subsubsec:estimated_slope_corrections_probabilities}, which quantifies the expected slope under the assumption of population-level reliability while accounting for finite-ensemble effects, without using any information about verifying observations or observed frequencies.
    Top panels show the number of cases in each probability bin.
    }
    \label{fig:tercile_reliability_diagrams_ecmwf}
\end{figure}

Figure~\ref{fig:tercile_reliability_diagrams_ecmwf} shows three complementary diagnostics of reliability for the middle and upper terciles.
First, binned averages of predicted probabilities and observed frequencies are presented, where forecast cases are stratified into 11 probability bins for each tercile.
Second, the observed frequency is regressed on the ensemble-based forecast probability across all cases, yielding an empirical slope that summarizes the forecast-to-observation relationship for each tercile.
Third, the slope is estimated using the ensemble-based estimator derived in Section~\ref{subsubsec:estimated_slope_corrections_probabilities}, which provides the theoretically expected slope under the assumption of population-level reliability while accounting for finite-ensemble sampling effects. This theoretical slope is constructed solely from the ensemble data, without any information about the verifying observations or the observed frequencies.

The reliability diagrams show that the empirical regression line closely follows the binned observed frequencies, indicating that the relationship between forecast probabilities and observed frequencies is well captured by a linear model.
However, the empirical slopes are substantially attenuated relative to the 1-to-1 line.
This attenuation is stronger for the middle tercile (slope = 0.128) than for the upper tercile (slope = 0.592).
Taken at face value, and without accounting for finite-ensemble effects, these results would suggest pronounced reliability deficiencies, particularly for the middle tercile.

In contrast, the ensemble-based slope estimator yields expected slopes of 0.128 and 0.586 for the middle and upper terciles, respectively.
These values quantify the degree of slope attenuation that is expected solely due to finite-ensemble sampling noise under the assumption of population-level reliability.
The close agreement between the empirical slopes and the theoretically expected values indicates that the observed attenuation is fully consistent with finite-ensemble effects, rather than reflecting genuine deficiencies in forecast reliability.
The stronger attenuation for the middle tercile is consistent with the smaller variability of its population probabilities across cases, as illustrated in Fig.~\ref{fig:tercile_reliability_diagrams}a.

What would change if the ensemble size were increased to 100 members?
To address this question, we analyse a small set of non-operational ECMWF re-forecasts with 100 ensemble members for the central tercile (see supplementary Fig.~S6).
This dataset comprises only 20 start dates, covering the period 2001-2020 with initializations on 1 February, so that sampling uncertainty associated with the limited number of cases is substantially larger than for the operational re-forecast dataset, which comprises 1220 start dates with 10 ensemble members.
To maintain comparability with the operational re-forecasts, results are shown both for a subset of 10 ensemble members and for the full 100-member ensemble, thereby isolating the effect of ensemble size.

Despite the increased (random) sampling uncertainty, a systematic reduction in slope attenuation when increasing the ensemble size from 10 to 100 members is clearly evident.
Both empirical and theoretically expected slopes lie substantially closer to the 1-to-1 relationship for the 100-member ensemble than for the 10-member ensemble, consistent with the expected reduction of slope attenuation with increasing ensemble size.
Moreover, the empirical slope remains in good agreement with its theoretically expected counterpart.
Differences between terciles persist, with the middle tercile continuing to exhibit stronger attenuation than the upper tercile (not shown).
Taken together, these results provide additional support for the interpretation that the observed deviations from the 1-to-1 line primarily reflect finite-ensemble sampling effects rather than genuine reliability deficiencies.

\section{Discussion}
\label{sec:discussion}

\subsubsection*{Relation to previous studies}

\textcite{Richardson2001} demonstrated that finite ensemble size leads to a systematic
flattening of probability-based reliability diagrams, even for perfectly reliable
forecast systems.
For a prescribed distribution of forecast probabilities across cases, they derived
analytical expressions for the expected observed frequencies in each probability bin
as a function of ensemble size.
Their analysis recovers the full conditional relationship between ensemble-based
probabilities and observed frequencies, but relies on explicit distributional
assumptions about how the population probabilities vary across cases, specifically
assuming a beta distribution.
As a consequence, the resulting expressions depend on unknown shape parameters that
must be specified or estimated.
In contrast, the present study focuses on the slope of the conditional relationship,
which captures the dominant effect of finite-ensemble sampling on reliability
diagrams.
By isolating this slope attenuation, we obtain analytical expressions and practical
estimators that depend only on ensemble-derived quantities and do not require
assumptions about the distribution of population probabilities across cases.
This makes the proposed correction directly applicable in operational settings,
where the variability of forecast probabilities is generally unknown and difficult
to model parametrically.

\textcite{ruppSpreadversuserrorFrameworkReliably2025} showed that finite ensemble size leads to a systematic flattening of the spread-error relationship and discussed the reason qualitatively.
Here, we provided a more general analytical framework that encompasses not only the spread-error relationship but also conditional verification of the ensemble mean and probability-based reliability diagrams.
Moreover, we derived practical estimators for the slope attenuation that can be computed directly from ensemble data without knowledge of population parameters.
Our synthetic data experiments closely follow their toy model analysis.

\subsubsection*{Comparison of forecast systems of different ensemble sizes}

When comparing conditional reliability diagnostics across different forecasting systems with varying ensemble sizes, it is crucial to account for the finite-ensemble slope attenuation effect identified in this study. Differences in diagnosed slopes may not solely reflect genuine differences in forecast reliability, but can also arise from variations in ensemble size and the associated sampling noise. Therefore, when interpreting such comparisons, it is advisable to apply the derived slope correction factors or estimators to ensure that observed differences are not confounded by finite-ensemble effects.

This issue is particularly relevant when evaluating conditional reliability for re-forecasts (or ``hindcasts''), which are widely used in sub-seasonal and seasonal prediction to assess forecast performance and calibrate forecast systems.
Conditional reliability diagnostics typically rely on re-forecast datasets because they span long periods, but the re-forecast ensemble size may differ from that of real-time forecasts. This is, for example, the case for ECMWF sub-seasonal forecasts, where the current operational re-forecast ensemble size is 11 members, while the real-time forecast ensemble size is 101 members.
When transferring conditional reliability diagnostics to real-time systems, it is therefore important to account for the impact of ensemble size on slope attenuation.
Otherwise, the uncorrected quantitative relationships between ensemble-derived quantities and their corresponding observed counterparts -- such as ensemble mean and expected mean, ensemble spread and expected error, or forecast probabilities and event frequencies -- will not be representative of the real-time forecast system.

In this study, we focused on the expected slope that a perfectly reliable forecast system would exhibit for the given ensemble size. Ongoing work aims to extend this framework to infer the slope that would be expected for any different ensemble size, by adjusting the noise variance terms in the derived expressions.
This would, for example, allow estimation of the expected slope for real-time forecasts based on slopes diagnosed from re-forecasts, thereby providing more appropriate guidance for users interested in the expected mean, error, or event likelihood associated with a given real-time forecast instance.

\subsubsection*{Reliability assessment versus practical forecast usefulness}

From a model development perspective, the results presented here show that conditional reliability can be meaningfully assessed even for relatively small ensembles, provided that finite-ensemble slope attenuation is properly accounted for.

From a user perspective, however, ensemble size remains crucial, but for a different reason.
Users are interested in probabilities that meaningfully depart from climatology (``sharpness'') and reliably translate into different expected outcomes (``resolution'').
While the total variability of issued probabilities decreases with increasing ensemble size, because sampling noise is reduced, the variability that is informative increases.
In small ensembles, a large fraction of the variability in issued probabilities is due to sampling noise, which inflates the apparent spread of probabilities without leading to corresponding differences in observed frequencies.
As ensemble size increases, this noisy variability is suppressed, but the remaining variability more faithfully reflects differences in the underlying population probabilities.

This distinction can be illustrated with concrete numbers for the central tercile shown in Section~\ref{subsec:reliability_diagrams_ecmwf}.
When using 10 ensemble members, 90\% of forecasts issue probabilities between 10\% and 60\%, suggesting a wide apparent range of forecast probabilities.
However, the empirical reliability slope is only 0.11, so that this wide range of issued probabilities translates into reliably distinguishable observed frequencies of only 27\% to 34\%.
In other words, much of the apparent probability variability arises from sampling noise and does not correspond to meaningful differences in expected outcomes.

When the ensemble size is increased to 100 members, the variability of issued probabilities is reduced, with 90\% of forecasts issuing probabilities between 19\% and 41\%.
At the same time, the empirical reliability slope increases to about 0.70.
As a result, this narrower range of issued probabilities translates into a wider and more informative range of reliably expected event frequencies, from 23\% to 36\%.
Thus, while reliability can be meaningfully assessed even for small ensemble sizes, larger ensembles provide substantially greater discrimination in terms of reliably different outcomes.

Note that in this study we use the term ``reliability'' to refer to a property of the forecast system: ensemble members and the verifying observation can be regarded as independent random draws from the same population distribution. As shown above, however, the issued probabilities of a perfectly reliable forecast system with finite ensemble size do not, in general, exactly match observed frequencies. One might therefore be tempted to call such probabilities ``unreliable'' in a practical sense. We avoid this terminology, since it would require an additional distinction between the ``reliability of the forecast system'' and the ``reliability of the issued probabilities''.

\subsubsection*{Assumptions and interpretation}

The analytical computation of slope attenuation assumes that ensemble members are exchangeable and can be interpreted as independent Monte Carlo draws from a case-specific population distribution.
If the verifying observation is also an independent draw from that same distribution, then the forecast system is perfectly reliable by definition.
In that case, conditional reliability diagnostics yield a slope of one in the limit of infinite ensemble size.

Accordingly, the derived slopes for the mean-, variance-, and probability-based diagnostics should be interpreted as the slopes expected from a perfectly reliable forecast system at finite ensemble size.
This provides a benchmark against which empirical slopes, estimated from forecast-observation data, can be compared.
If the empirical and expected slopes are consistent within sampling uncertainty arising from a finite number of forecast cases, then the observed attenuation can be explained by finite-ensemble effects alone and does not provide evidence of genuine reliability deficiencies.

In practice, the ability to reject perfect reliability can also depend on ensemble size.
For a fixed number of forecast cases and a fixed departure from perfect reliability at the population level, smaller ensembles induce stronger attenuation and therefore reduce the separation between the diagnosed slope and the slope expected under perfect reliability, which might make such departures harder to detect.

Importantly, a slope of one is a necessary but not sufficient condition for perfect reliability, since other forms of model error may not be detected by slope-based diagnostics.

Our derivations hold exactly only in the limit of an infinite number of forecast cases, so that empirical statistics converge to their expected values (see next paragraph).

The derived slope attenuation expressions do not depend on the specific form of the population distributions.
The synthetic data experiments presented here use Gaussian population distributions for simplicity, but supplementary Fig.~S5 shows that the same conclusions also hold for a strongly skewed normal population distribution.

\subsubsection*{Sampling uncertainty due to finite number of cases}

Beyond the finite-ensemble effects that are the focus of this study, empirical estimation of conditional slopes is also affected by sampling uncertainty arising from a finite number of forecast cases. This source of uncertainty is conceptually distinct from finite-ensemble sampling noise. While finite ensemble size leads to a systematic attenuation of conditional slopes, a limited number of forecast cases does not introduce a systematic bias but instead adds random noise to the slope estimates.

Supplementary Figure~S4 illustrates the impact of finite case numbers on slope estimates using synthetic data experiments. The results presented are identical to those in Fig.~\ref{fig:slope_estimators}, but now using only 1,000 forecast cases instead of 200,000. The reduced sample size leads to substantial sampling uncertainty, which is most pronounced for the empirical slopes computed directly from observations. In contrast, the ideal and estimated slopes, which rely on ensemble-derived statistics, exhibit considerably less variability. This is because the ensemble-based estimators effectively utilize information from all ensemble members, whereas empirical slopes depend solely on the single verifying observation per case.
First and foremost, this highlights the importance of quantifying sampling uncertainty when interpreting conditional reliability diagnostics. Second, it limits the practical applicability of the derived slope correction factors. Even if the expected slope attenuation due to finite ensemble size can be accurately estimated, the presence of substantial sampling uncertainty in empirical slopes may obscure genuine differences in reliability. Therefore, caution is warranted when applying slope corrections in settings with limited case numbers.

In practice, the number of forecast cases available for a given location and season
is often limited, frequently well below 1,000.
Although pooling data from larger spatial domains or multiple seasons can increase the effective sample size and reduce sampling uncertainty, such aggregation can also mask the structure of the conditional relationships of interest \parencite[cf.][]{hamillMeasuringForecastSkill2006}.
We therefore recommend first estimating conditional slopes separately for each location and season, and only then averaging the resulting slope estimates over space and time.
Pooling all data points into a single regression can artificially inflate the case-to-case variability of the population parameters due to spatial and seasonal differences, thereby emphasizing climatological contrasts rather than the intrinsic flow-dependent behaviour at a given location and season.

\subsubsection*{Role of model error}

Model error introduces additional complexity into the interpretation of conditional reliability diagnostics. Importantly, deficiencies in the representation of mean quantities need not coincide with deficiencies in the representation of their variability. For example, a forecast model may exhibit a biased climatological mean while still accurately representing the variability of the population mean across cases, resulting in reliable conditional relationships despite the bias. Conversely, a model may have an unbiased mean state but underestimate the variability of the population mean, leading to attenuated conditional slopes that reflect genuine reliability deficiencies rather than finite-ensemble effects.

Another class of model error arises when the response of population parameters to changes in the underlying state of the system is misrepresented.
In our synthetic experiments, we assumed that there is one population distribution per case, which is shared between the model and the observations.
However, in real-world forecasting systems, the model distribution may differ from the (hypothetical) true observational population distribution even if the ensemble size was infinite.
This is therefore distinct from the finite-ensemble attenuation discussed above and instead reflects a genuine deviation from reliability due to model bias.
For instance, the model may underestimate the sensitivity of population parameters to predictable signals.
In the presence of a predictable signal such as El Niño, the population mean may shift substantially, while the model underestimates the amplitude of this shift.
In this case, the slopes of conditional relationships exhibit slope steepening rather than attenuation, as previously discussed for probability-based reliability diagrams \parencite{strommenRelationshipReliabilityDiagrams2023} and for spread-error relationships \parencite{ruppSpreadversuserrorFrameworkReliably2025}.
This is because a slight positive anomaly in the forecast may correspond to a larger anomaly in the observed quantity. The same logic applies to the ensemble mean, the ensemble spread, and event probabilities, yielding slopes greater than one.
Such effects would also arise within the framework presented here, although we have not specifically explored them in this study.

\subsubsection*{Drivers of population variability}

A central result of this study is that the magnitude of slope attenuation in conditional reliability diagnostics depends on the case-to-case variability of the underlying population parameters.
This was qualitatively outlined in \textcite{ruppSpreadversuserrorFrameworkReliably2025}.
In practice, such variability can arise from a wide range of mechanisms.
Shifts of the population distribution may be driven by predictable large-scale signals, such as seasonal cycles, long-term trends (e.g., due to climate change), or teleconnections including ENSO.
Changes in the spread of the population distribution may reflect variations in the predictability of the system, which can also be modulated by seasonal cycles, trends, or teleconnections.
Additional higher moments of the population distribution, such as skewness or kurtosis, may also vary across cases due to changes in the underlying dynamics or external forcings.

The sources of population variability discussed above also arise from methodological choices in the construction of conditional reliability diagnostics.
When diagnostics are intended to assess flow-dependent variations in mean, ensemble spread, or forecast probabilities, they may be evaluated for fixed locations and seasons.
In contrast, however, reliability diagrams are sometimes constructed by pooling forecasts from different locations and seasons.
Such pooling typically introduces substantial additional population variability, as the underlying population parameters may differ systematically across space and time.
For example, ensemble forecasts of near-surface temperature generally exhibit larger spread over land, at higher latitudes, and during winter, compared to over the ocean, at lower latitudes, or during summer.
When forecasts from these heterogeneous regimes are pooled, the slope attenuation effect is reduced through inter-seasonal and inter-spatial inflation of population variability.
However, in this setting, conditional reliability diagnostics may primarily reflect the model's ability to capture seasonal cycles and climatological spatial contrasts, rather than flow-dependent variations within a given regime.
In the example of near-surface temperature spread, the largest-spread bins are then dominated by winter forecasts at high latitudes over land, while the smallest-spread bins are dominated by summer forecasts at low latitudes over the ocean.

\subsubsection*{Conditioning on unrelated quantities}

Finally, it is important to note that the systematic attenuation effects discussed in this study arise specifically when conditioning is performed on noisy ensemble-derived quantities. Conditioning on external or unrelated variables, such as observed indices describing the ENSO state or the strength of the stratospheric polar vortex at initialization, does not introduce the same bias. In such cases, ensemble-derived sampling noise does not enter the conditioning variable, and conditional diagnostics are not subject to finite-ensemble slope attenuation. This distinction underscores that the bias identified here is not a generic property of conditional verification, but a consequence of conditioning on predictors that themselves are estimated from finite ensembles.

\section{Conclusions}
\label{sec:conclusions}

Conditional reliability diagnostics such as reliability diagrams and spread-error relationships are widely used to assess ensemble forecasts.
In ensemble mean reliability diagrams, forecast ensemble means are stratified and compared to observed means.
In ensemble spread reliability diagrams, forecast ensemble spreads are stratified and compared to observed forecast errors. 
In probability-based reliability diagrams, forecast event probabilities are stratified and compared to observed event frequencies.
It has been common practice to interpret the slope of the linear fit in these diagnostics as a measure of reliability, with the 1-to-1 line representing the benchmark of perfect reliability.
In this study, however, we showed that this benchmark must be qualified for finite ensembles: even for a perfectly reliable forecast system, the relationship between ensemble-derived statistics and their corresponding observed counterparts can deviate substantially from the 1-to-1 line.
This deviation arises from sampling noise associated with finite ensemble size and manifests itself as systematic slope attenuation.

We explained the origin of this effect and derived analytical expressions for the expected slope attenuation.
The attenuation is stronger when the ensemble size is smaller and when the variability of the underlying population parameters across cases is weaker.
We further derived estimators that allow the expected slope attenuation to be estimated from finite ensemble data alone, without observational data, without knowledge of the underlying population parameters, and without distributional assumptions.
Interpreting attenuated conditional slopes as forecast deficiencies without accounting for finite-ensemble effects can therefore be misleading.

More broadly, the results show that finite ensemble size affects not only the verification of forecasts, but also their interpretation.
Even for a perfectly reliable forecast system, a finite-ensemble probability \(p\) does not imply an expected event frequency of exactly \(p\).
Rather, finite-ensemble probabilities are systematically shifted away from the climatological mean: probabilities above climatology are on average associated with lower observed frequencies, and probabilities below climatology with higher observed frequencies.
This mismatch is an intrinsic consequence of finite-ensemble sampling, not a sign of model failure. Recognizing it is therefore essential both for evaluating and for using ensemble forecasts.

\subsubsection*{Acknowledgments}
This work builds on the foundation laid by previous joint work with Philip Rupp, whose contributions to the development of the underlying framework and the synthetic data experiments are gratefully acknowledged.
We particularly thank Llorenç Ledó for providing very valuable feedback on the manuscript.
We also thank Daniel Befort, Martin Leutbecher, Frédéric Vitart, Linus Magnusson, Thomas Haiden and Steffen Tietsche for insightful discussions that helped shape the ideas presented in this study.
This work was supported by the European Union's Horizon Europe research and innovation programme under grant agreement No. 101081568: ``Arctic Cross-Copernicus forecast products for sea Ice and iceBERGs'' (ACCIBERG).

\subsubsection*{Data and code availability}
Operational ECMWF sub-seasonal re-forecast data are available from the ECMWF data portal (\url{https://apps.ecmwf.int/datasets/data/s2s}).
Python code to reproduce the synthetic data experiments and evaluate the slope estimators is available at [to be added during review process].

\subsubsection*{Author contributions}
JS conceived the study, developed the analytical framework, performed the analyses, and drafted the manuscript.
CR contributed to the formulation of the conceptual research questions, to the interpretation of the results, laid the foundation for extending the framework to probability-based reliability diagrams, and provided feedback that improved the manuscript.

\subsubsection*{Competing interests}
The authors declare no competing interests.

% \printbibliography
\bibliographystyle{rss.bst}
\bibliography{references.bib}

\end{document}

%% file: style.tex
\usepackage[utf8]{inputenc}
\usepackage[english]{babel}

\usepackage{amsmath, amssymb}
\usepackage{siunitx}
\usepackage[acronym]{glossaries}
\newacronym{SPP}{SPP}{Stochastically Perturbed Parametrisation}
\newacronym{fCRPSS}{fCRPSS}{fair Continuous Ranked Probability Skill Score}
\newacronym{t2m}{T2m}{2-metre temperature}
\newacronym{ECMWF}{ECMWF}{European Centre for Medium-Range Weather Forecasts}
\newacronym{IFS}{IFS}{Integrated Forecasting System}
\newacronym{SPS}{SPS}{Spatial Probability Score}
\usepackage{csquotes}
\usepackage{natbib}
\let\parencite\citep
\let\textcite\citet

\usepackage{graphicx}
\graphicspath{{./../figures}}
\usepackage{caption}
\usepackage{subcaption}
\usepackage{float}

\usepackage{booktabs}
\usepackage{multirow}

\usepackage[dvipsnames]{xcolor}
\usepackage{xr-hyper} % allow to refer to supplement labels
\usepackage[hidelinks]{hyperref}

\usepackage[a4paper,margin=2.5cm]{geometry}

\usepackage{array}
\newcolumntype{L}[1]{>{\raggedright\hangindent=1em\arraybackslash}p{#1}}

\usepackage[left,modulo]{lineno}  % "left" for numbers on left margin

\usepackage{authblk}

\usepackage{orcidlink}

\usepackage{fancyhdr}
\usepackage{lastpage}